\documentclass[article, twocolumn]{revtex4}
\usepackage[utf8]{inputenc}
\usepackage[english]{babel}
\usepackage{graphicx}
\usepackage{color}
\usepackage{bm}
\usepackage{amssymb}
\usepackage[intlimits]{amsmath}
\usepackage{amsmath}
\usepackage{tabu}
\usepackage{longtable}

\begin{document}

\title{The amplitudes and the structure of the charge density wave in YBCO}
\author{Y.~A.~Kharkov}
\author{O.~P.~Sushkov}
\email{sushkov@unsw.edu.au}
\affiliation{School of Physics, University of New South Wales, Sydney 2052,
Australia}

\begin{abstract}
We find unknown $s$- and $d$-wave
amplitudes of the recently discovered
charge density wave (CDW) in underdoped cuprates.
To do so we perform a combined analysis of experimental data
for ortho-II YBa$_2$Cu$_3$O$_{y}$.
The analysis includes data on nuclear magnetic resonance, 
resonant inelastic X-ray scattering, and 
hard X-ray diffraction.
The amplitude of doping modulation found in our analysis
is $3.5\cdot 10^{-3}$ in a low magnetic field and $T=60$K,
 the amplitude is $6.5\cdot 10^{-3}$ in a magnetic field of 30T and $T=1.3$K. 
The values are in units of elementary charge per unit cell of a CuO$_2$ plane.
We show that the data rule out a checkerboard pattern, and we also show
that the data might rule out mechanisms of the CDW which do not include phonons.
\end{abstract}

\maketitle

The recent discovery of the charge density wave (CDW) in YBCO and other cuprates gave 
a new twist to physics of high-T$_c$ superconductivity. Existence of a new charge 
ordered phase has been reported in bulk sensitive nuclear magnetic  resonance (NMR) 
measurements~\cite{Wu2011, Wu2013, Wu2015}, resonant inelastic X-ray 
scattering (RIXS)~\cite{Ghiringhelli2012, Achkar2012, Blackburn2013}, 
resonant X-ray scattering \cite{Comin2014} and hard X-ray diffraction (XRD)~\cite{Chang2012}.
Additional non-direct evidence comes from measurements of ultrasound speed~\cite{LeBeouf2012} 
and Kerr rotation angle~\cite{Xia2008}.

While the microscopic mechanism of the CDW and its relation to
superconductivity remains an enigma,
there are several firmly established facts listed below, 
here we specifically refer to YBCO. (i) The CDW state arises
in the underdoped regime within the doping range $0.08 \leq p \leq 0.13$.
(ii) The onset temperature of CDW at doping $p\sim0.1$
is $T_{CDW} \approx 150$ K, which is between the
pseudogap temperature $T^*$ and the superconducting temperature $T_c$, $T_c<T_{CDW}<T^*$.
(iii) The CDW ``competes'' with superconductivity, the CDW amplitude is suppressed
at $T < T_c$. Probably due to this reason the CDW amplitude at $T < T_c$
is enhanced by a magnetic field that suppresses superconductivity.
 (iv) The CDW  wave-vector is  directed along the CuO link in the CuO$_2$ plane. 
(v) The wave-vector $Q\approx 0.31$ r.l.u. only very weakly depends on doping.
(vi) The CDW is essentially two-dimensional in low magnetic fields, the correlation length 
in the $c$-direction is about one lattice spacing, while the in-plane
correlation length is $\xi_{a,b} \sim 20$  lattice spacings. (vii) In high magnetic fields ($B>15$ T)  and low temperatures ($T<50$ K) the CDW exhibits three-dimensional correlations with the correlation length in the $c$-direction $\xi_c \sim 5$ lattice spacings \cite{Gerber2015, Chang2015}.
(viii) Ionic displacements in the CDW are about $10^{-3}\AA$~\cite{Forgan2015}.

In spite of numerous experimental and theoretical works, there are 
two major unsolved problems in the phenomenology of the CDW.
(i) The amplitude of the electron density modulation remains undetermined.
(ii) The intracell spatial charge pattern is unclear,
while there are indications from RIXS~\cite{Comin2015} and from 
scanning tunneling microscopy~\cite{Hamidian2015}
that the pattern is a combination of $s$- and $d$-waves.
The major goal of the present work is to resolve the open problems.
We stress that in the present paper  we  perform combined analysis of 
experimental data to resolve the problems of the phenomenology,
but we do not build a microscopic model of the CDW.
While we rely on various data, the most important information in this
respect comes from NMR. In particular we use the ortho-II YBCO  data.
Ortho-II YBCO (doping $p\approx 0.11$) is the least disordered underdoped 
cuprate and hence it has the narrowest NMR lines. Development of the CDW with decreasing of 
temperature leads to 
the broadening of the quadrupole satellites in the NMR spectrum ~\cite{Wu2011, Wu2013, Wu2015}.  
Below we refer the quadrupole satellites as NQR
lines. Quite often the term "NQR" implies zero magnetic field 
measurements. We stress that it is not true in our case,
NQR here means quadrupole satellites of NMR lines. The broadening is directly proportional to the CDW amplitude with the 
coefficients determined in Ref.~\cite{Haase2004}.
So, one can find the CDW amplitude and this is the idea of the present analysis.
Moreover, combining the data on copper and oxygen NMR we deduce the CDW intracell pattern within the CuO$_2$ plane.

The second goal of the present work is ``partially theoretical''.
Based on
the phonon softening data~\cite{LeTacon2014} we are able
to separate between two broad classes of possible mechanisms responsible 
for the formation of the CDW.
(i) In the first class the CDW is driven purely by strongly correlated
electrons which generate the charge wave. In this case phonons and the lattice 
are only spectators which follow  electrons.
(ii) In the second class, which we call ``the Peierls/Kohn'' scenario, both
electrons and phonons are involved in the CDW development on equal footing.
We argue that the phonon softening data ~\cite{LeTacon2014}  potentially supports the second scenario.

\begin{figure}[h]
\includegraphics[scale=0.13]{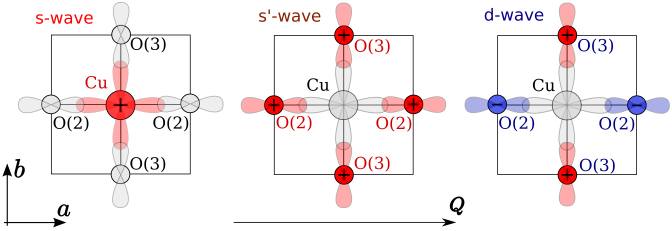}
\caption{Intra-unit cell patterns of the CDW directed along the a-axis,
${\bm Q}=(Q,0)$: $s$-wave, $s'$-wave, and 
$d$-wave. Positive and negative excess charge variations are shown in red and blue
respectively.
 }\label{F1}
\end{figure}

The CDW implies modulation of electron charge density on copper and oxygen 
sites in the CuO$_2$-planes. Our notations correspond to the orthorhombic YBCO,
the axis c is orthogonal to the CuO$_2$-plane, the in-plane axes $a$ and $b$
are directed perpendicular and parallel to the oxygen chains, respectively.
Usually the CDW is described in terms of $s$-, $s'$-, and $d$-wave components
with amplitudes $A_s$, $A_{s'}$, and $A_{d}$,
see e.g. Refs.~\cite{Comin2015, Fujita2014, Hamidian2015}. 
The $s$-wave component corresponds to the modulation of the population of Cu $3d_{x^2-y^2}$
orbitals, and $s'$- and $d$-wave components correspond to the modulation
of the populations of oxygen $2p_{\sigma}$ orbitals:
\begin{eqnarray}
\label{para}
&&\delta n_d = A_s \cos[{\bm Q}\cdot{\bm r}+\phi_s], \\ 
&&\delta n_{px} = A_{s'}\cos[{\bm Q}\cdot{\bm r}+\phi_{s'}] 
+ A_{d}\cos[{\bm Q}\cdot{\bm r}+\phi_d],
 \nonumber\\ 
&&\delta n_{py} = A_{s'}\cos[{\bm Q}\cdot{\bm r}+\phi_{s'}] 
- A_{d}\cos[{\bm Q}\cdot{\bm r}+\phi_d]. \nonumber
\end{eqnarray}
Here $\bm Q$ is the wave vector of the CDW,
 directed along $a$ or $b$ crystal axis [${\bm Q}=(Q,0)$ or ${\bm Q}=(0,Q)$]
 and
$\phi_s$, $\phi_{s'}$, $\phi_d$ are the phases of $s$-, $s'$- and $d$-waves. The subscripts 
``$x$'' and ``$y$'' in Eq.~(\ref{para}) indicate different oxygen sites within 
the CuO$_2$-plane unit cell.
The standard nomenclature of the oxygen sites in YBCO is O(2) and O(3).
The O(2) 2p$_\sigma$-orbital is parallel to the axis ``$a$'', and
the O(3) 2p$_\sigma$-orbital is parallel to the axis ``$b$'', see 
Fig.~\ref{F1}.
For the CDW wave-vector $\bm Q$ directed along the $a$-axis,  the ``$x$''-site
is O(2) and the ``$y$''-site is O(3) as shown in Fig.~\ref{F1}.
In the same figure we indicate excess charge corresponding to
$s$, $s'$-, and  $d$-waves.
For ${\bm Q}$ orientated along the axis ``$b$'' the ``$x$''-site is O(3),
and the ``$y$''-site is O(2).

According to the analysis~\cite{Haase2004} the NQR frequency  of a particular
$^{17}$O nucleus is proportional to the local hole density $n_p$ at this site,
and of course it depends on the orientation of the magnetic field with respect
to the oxygen p-orbital,
\begin{eqnarray}
\label{Onu}
&&B \perp   2p_{\sigma}: \
f_{O\perp}\approx 1.23 MHz \times n_p +C_1, \nonumber\\
&& B \ || \  2p_{\sigma}: \ 
f_{O||}\approx 2.45 MHz \times n_p +C_2,
\end{eqnarray}
where $B$ is the external magnetic field of NMR.
Constants $C_1$ and $C_2$ are due to other ions in the lattice;
generally they depend on the position of the oxygen ion in the lattice.
Typical values of these constants are: $C_1 \sim 0.2$~MHz, $C_2 \sim 0.5$~MHz.
According to the same analysis ~\cite{Haase2004}  
the $^{63}$Cu NQR frequency is proportional to the local hole density $n_d$ at 
the Cu site and also $n_p$ at the adjacent oxygen sites,
\begin{eqnarray}
\label{Cunu} 
\text{Cu}^{63} (B\, || \,{ c}): \ f&\approx& 94.3 MHz \times n_d \\
            & -&11  MHz \times[4-(n_{pa}+n_{pb})]+C_3 .\nonumber
\end{eqnarray}
Here the ``ion-related'' constant $C_3 \sim -6$ MHz.

There are two mechanisms for the
position dependent variation of the NQR frequency which are related to the CDW,  
(i) a variation of the local densities $n_d, n_p$, (ii) a variation of the ions' positions. The position dependent frequency variation leads to the
observed inhomogeneous broadening of the NQR line.
Let us show that the mechanism (ii) is negligible.
Only in-plane displacements of ions contribute to (ii) in the first order
in the ion displacement.
The magnitudes of the relative in-plane displacements of Cu and O ions are 
$\delta r/r \lesssim  10^{-3}$~\cite{Forgan2015}, where $r\approx 2\AA$
is the Cu-O distance. Hence we can expect a lattice-related
variation of e.g. oxygen $f_{\perp}$ at the level 
$\delta f_{\perp} \sim C_1\delta r/r \sim 0.2$ kHz.
This is two orders of magnitude smaller than the CDW related broadening 
$\sim 10$ kHz observed experimentally.
For copper nuclei the expected ion-related broadening
comes mainly from the $11\times 4$MHz term in ~(\ref{Cunu}),  
$\delta f \sim \delta r/r\times 44 \text{ MHz }\sim 0.04$ MHz. Again, this is much smaller
than the observed broadening $\sim 1$ MHz.
These estimates demonstrate that one can neglect the 
contribution of the lattice distortion in the NQR broadening.
Therefore, below we consider only the broadening mechanism (i) related to 
variation of hole densities.

Any compound has an intrinsic quenched disorder.
The disorder is responsible for the NQR line widths at $T>T_{CDW}$.

\begin{figure}[h]
\includegraphics[scale=0.07]{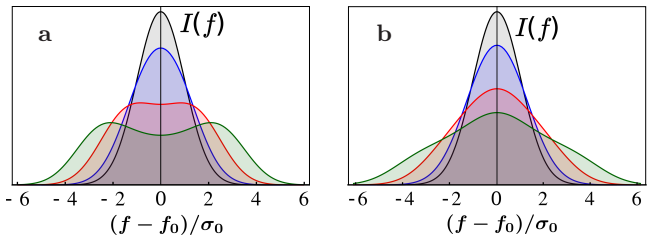}
\begin{picture}(0,0)
         \put(-233,73){\textbf{a}}
     \put(-105,73){\textbf{b}}
\end{picture}
\caption{\textbf{(a)} The NQR lineshapes for the stripe-like CDW ~(\ref{anu}).
\textbf{(b)} The NQR lineshapes for the  checkerboard CDW ~(\ref{anu2}).
Both \textbf{(a)} and \textbf{(b)} show the lines for four different values of the CDW amplitude $A$
with respect to the intrinsic broadening, $A/\sigma_0=0,1,2,3$.
}
\label{F2}
\end{figure}

The experimental NQR lines in a ``weak magnetic field'', $B = 12-15$T, are practically symmetric, the analysis of the NQR lines and the 
corresponding values of full widths at half maximum (FWHM) are presented in Refs. \cite{Wu2013, Wu2015}.
However, the experimental NQR lines in a ``strong magnetic field'' \cite{Wu2011},
$B \approx 30$T, are somewhat asymmetric due to various reasons. The asymmetry brings a small additional
uncertainty in the analysis. The ``strong field'' data is less 
detailed than the ``weak field'' data and therefore the additional uncertainty
is completely negligible, the ``strong field'' NQR widths are given in Ref. \cite{Wu2011}.
Hereafter we assume simple Gaussian lines, 
$I(f)\propto \exp[-(f-f_0)^2/2\sigma_0^2]$, where $f_0$
is the center of the NQR line, $\sigma_0$ corresponds to the 
intrinsic disorder-related width.
 At $T < T_{CDW}$ the line shape is changed to
\begin{equation}
\label{ls}
I(f) \propto \left\langle \exp\left\{-\frac{[f-f_0-\delta f({\bf r})]^2}
{2\sigma_0^2}\right\}\right\rangle\ ,
\end{equation}
where $\delta f({\bm r})$, 
\begin{equation}
\label{anu}
\delta f({\bm r})=A\cos[({\bm Q}\cdot{\bm r})+\phi]\ 
\end{equation}
follows from Eqs.~(\ref{Onu}),~(\ref{Cunu}),~(\ref{para}).
In particular, in MHz
\begin{eqnarray}
\label{anu1}
\delta f_{O\perp}=1.23\left\{A_{s'}\cos[{\bm Q}\cdot{\bm r}+\phi_{s'}] 
\pm A_{d}\cos[{\bm Q}\cdot{\bm r}+\phi_d]\right\}, \nonumber\\
\delta f_{O ||}=2.45\left\{A_{s'}\cos[{\bm Q}\cdot{\bm r}+\phi_{s'}] 
\pm A_{d}\cos[{\bm Q}\cdot{\bm r}+\phi_d]\right\}, \nonumber\\
\delta f_{Cu}=94.3 A_s \cos[{\bm Q}\cdot{\bm r}+\phi_s]
+22 A_{s'}\cos[{\bm Q}\cdot{\bm r}+\phi_{s'}]. \
\end{eqnarray}
The averaging in Eq.~(\ref{ls}), $\left\langle ....\right\rangle$, is 
performed over the position ${\bf r}$ of a given ion (Cu or O) in 
the CuO plane.
A simulation of $I(f)$ in Eq.~(\ref{ls}) with $\delta f$ from ~(\ref{anu})
is straightforward, the results for several values of the ratio $A/\sigma_0$
are presented in Fig.~\ref{F2}a. 
The CDW leads to the NQR line broadening and at larger amplitudes 
results in a distinctive double peak structure. 
For a comparison in the Panel b of Fig.~\ref{F2} we present the lineshapes
obtained with Eq.~(\ref{ls}) for the checkerboard density modulation,
\begin{equation}
\label{anu2}
\delta f({\bm r})=\frac{A}{\sqrt{2}}[\cos(Qr_a) + \cos(Qr_b)] \ .
\end{equation}
Obviously, the lineshapes in panels a and b of Fig.~\ref{F2} are very different.
The checkerboard pattern does not result in the  double peak structure
even at very large amplitudes.
NQR data~\cite{Wu2011,Wu2013} clearly indicate the double peak structure.
This is a fingerprint of the  stripe-like CDW.
Comparison of the experimental NQR lineshapes  with Panels b of Fig.~\ref{F2}
rules out the checkerboard scenario at large magnetic field, see also \cite{Comin2015stripe, Fine2016, Comin2016}. 

\begin{figure}
\includegraphics[scale=0.22]{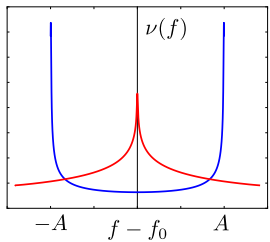}
\caption{``Density of states'' $\nu(f)$ for the stripe CDW (blue line) and for the checkerboard CDW (red line). The two singularities at $f-f_0 = \pm A$ in the case of the stripe-like CDW result in the double peak structure of the NQR lines.}\label{fig:density_of_states}
\end{figure}

The qualitative difference between the lineshapes corresponding
to the stripe and the checkerboard patterns is a ``density of states'' effect. 
Indeed, the NQR intensity (\ref{ls}) can be written in terms of the ``density of states''
\begin{eqnarray}
I(f)\propto \left\langle \ldots \right\rangle = \int dS \, (\ldots) = \int df \, \nu(f) (\ldots),
\end{eqnarray}
where $(\ldots)$ denotes the Gaussian exponent in (\ref{ls}), $dS$ is the element of area in the CuO$_2$ plane.
The ``density of states'' $\nu(f)=\int dS\, \delta (f - f_0 - \delta f(\bm r))$ in the case of the stripe-like CDW
(\ref{anu}) has two singularities, see Fig.~\ref{fig:density_of_states}, at  points $f-f_0 = \pm A$.
The singularities result in two peaks in the NQR spectrum.
On the other hand, in the case of the
checkerboard CDW (\ref{anu2}) the ``density of states'' $\nu(f)$ has
a single maximum  at $f = f_0$, see Fig.~\ref{fig:density_of_states}, leading to a single-peak
NQR lineshape. 
After averaging over the $(Q, 0)$ and $(0, Q)$ stripe domains
the double-peak NQR lineshape is intact. 
Of course this is true only because the size of the domains 
($\xi_{a,b} \sim 20-60 $ lattice spacings) is much larger then the period of the 
CDW ($2\pi /Q \approx 3.2$ lattice spacings).

Numerical integration (averaging) in (\ref{ls}) shows that the full NQR line width at half maximum can be approximated as
\begin{eqnarray}
\label{tlw}
&&\Gamma_0=\sqrt{8\ln 2}\sigma_0,\nonumber\\
&& \Gamma \approx \sqrt{\Gamma_0^2+4\ln 2A^2} \ , 
\ \ \ \ \ \ \ \text{if} \ \ \Gamma < \Gamma_0 ,
\nonumber\\
&&\Gamma \approx 1.2\sqrt{\Gamma_0^2+4\ln 2A^2} \ ,
\ \ \ \text{if} \  \ \Gamma > 2 \Gamma_0 . 
\end{eqnarray}
Note that this is the FWHM even for the non-Gaussian
line shape like that in Fig.~\ref{F2}a.
All the data we use below are in the regime $\Gamma \geq 2 \Gamma_0$.

\begin{figure}[h]
\vspace{10pt}
\includegraphics[scale=0.3]{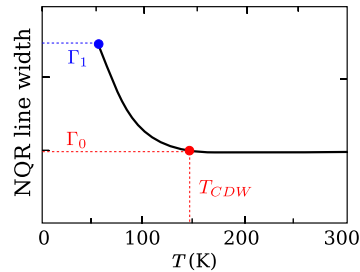}
\vspace{-10pt}
\caption{A schematic sketch of the temperature dependence of the NQR line width \cite{Wu2015}.
$\Gamma_0$ is the width at the temperature $T_{CDW}$, where the broadening starts to increase
with temperature lowering, and $\Gamma_1$ is the width at the lowest
available temperature.
 }
\label{F3}
\end{figure}

The typical dependence of the experimental NQR line width~\cite{Wu2015}
on temperature is sketched in Fig.~\ref{F3}. 
Hereafter we denote by $\Gamma_0$ the value of the width at the temperature, where the 
width starts to increase with  lowering of temperature, and by $\Gamma_1$ the 
value of the width at the lowest available temperature, as indicated in Fig.~\ref{F3}.
The increase of the width at low temperatures is due to the CDW development.
Comparing the data with Eqs.~(\ref{tlw}) we can find the CDW amplitudes.
There are two distinct Cu positions in ortho-II YBCO, Cu(2E) and Cu(2F),
that reside under the empty (E) and full( F) oxygen chains, respectively.
There are also three distinct in-plane oxygen positions, O(2), O(3F), and
O(3E). The O(2) $2p_{\sigma}$ orbital is oriented along the $a$-axis,
and the O(3F), O(3E) $2p_{\sigma}$ orbitals are  oriented along the $b$-axis 
(see Fig.~\ref{F1}).
O(3F) resides under the full chain and O(3E) resides under the empty
chain. The NQR broadening data for Cu(2E) and Cu(2F) are almost identical,
the same is true for O(3F) and O(3E).
Therefore, in our analysis we do not distinguish ``E'' and ``F''  and refer to them
as Cu(2) and  O(3) respectively.

It is worth noting that the NQR lines have been measured in NMR
experiments. Therefore, the actual line broadening is a combined effect
of the NQR broadening and the NMR broadening. The NMR broadening
comes from the magnetic Knight shift which is proportional to the charge 
density modulation. The  Knight shift broadening is itself an interesting
effect which can bring additional information about CDW.
However, our present analysis is aimed at NQR.
The structure of NMR satellites enables the subtraction of the
Knight shift effect from the data. The subtraction results in the
``rectified'' NQR line widths, which we use in our analysis.
The rectified NQR line widths obtained in Refs.~\cite{Wu2015,Julien}
for different ions and for different orientations of magnetic field
are listed in the second and third columns of Table~\ref{t1}.
In this case the magnetic field is  $B = 12-15$T, $\Gamma_0$ corresponds to
150K and $\Gamma_1$ corresponds to 60K.
\begin{table}[h!]
\begin{center}
\begin{tabular}{ | l | l  l  l | } 
 \hline
                  &   $\Gamma_0(kHz) $ & $\Gamma_1(kHz) $ &  $|A_s + 0.23 A_{s'}|$\\ 
 \hline
Cu(2), \ B$||$c & \ \ 230(30)       &   \ \ 460(80)  &\ \  2.0(0.5) \\ 
 \hline 
 & & & $\sqrt{A_{s'}^2 + A_d^2}$\\
 \hline
O(2), $2p_{\sigma}|| a$, B$||$a    & \ \  6.0(0.5)      & \ \ 16.0(1.3) & \ \  2.9(0.3)\\ 
 \hline
O(2), $2p_{\sigma}|| a$,   B$||$b    & \ \  6.0(0.6)     & \ \ 15.0(0.8) &\ \ 5.4(0.4)  \\ 
 \hline
O(2),  $2p_{\sigma}|| a$,   B$||$c ${+20^o}$ & \ \ 5.0(2.0) \ \  & \ \ 16.0(2.5) &\ \ 6.0(1.1) \\ 
 \hline
O(3),  $2p_{\sigma}|| b$,     B$||$a   & \ \ 6.0(1.0)    & \ \ 11.0(1.8) & \ \ 3.6(0.9) \\ 
 \hline
O(3), $2p_{\sigma}|| b$,     B$||$b  & \ \  5.0(2.0)     & \ \ 12.0(2.0) & \ \ 2.1(0.5)  \\ 
 \hline
O(3), $2p_{\sigma}|| b$,   B$||$c +20$^o$ & \ \  9.0(2.3) & \ \  18.0(2.3) &\ \ 6.1(1.4) \\
 \hline
\end{tabular}
\caption{ NQR data for ortho-II YBCO in magnetic field  $B = 12-15$T.
The line widths, $\Gamma_0=\Gamma_{T=150K}$
and $\Gamma_1=\Gamma_{T=60K}$, are measured for different
ions and for different orientations of the 
magnetic field~\cite{Wu2015,Julien}.
The last column displays the CDW amplitudes deduced from the
particular line. Figures in brackets represent crude estimates
of error bars.
}
\label{t1}
\end{center}
\end{table}
Figures in brackets represent crude estimates of error bars.

The CDW-induced broadening at oxygen sites comes from contributions of the $s'$- and $d$-waves, see Eq.~(\ref{anu1}).
RIXS and XRD data~\cite{Ghiringhelli2012, Achkar2012, Chang2012, Blackburn2013} suggest that the CDW state consists 
of equally probable domains with the one-dimensional CDW along $(Q,0)$ and $(0,Q)$ directions.
This means, that even if the phases of $s'$ and $d$-wave are locked in a domain, say $\phi_{s'}=\phi_d$, 
the $s'$-$d$ interference disappears from the oxygen broadening after averaging over orientations of the domains.
Hence, comparing Eqs.~(\ref{anu}) and ~(\ref{anu1}) we conclude that for the oxygen sites
$A=K\sqrt{A_{s'}^2+A_d^2}$ with the coefficient $K=1.23$MHz  or $K=2.45$MHz dependent on the orientation of the
magnetic field.
Using the experimental widths presented in Table~\ref{t1} together with Eq.~(\ref{tlw}) one finds
values of $\sqrt{A_{s'}^2+A_d^2}$ for each particular oxygen ion and orientation of the magnetic field.
The values with indicative error bars are listed in the last column of Table~\ref{t1}.
The average over the six different cases presented in the Table is
\begin{equation}
\label{o1}
\sqrt{A_{s'}^2+A_d^2}= 3.8 \cdot 10^{-3}\ .
\end{equation}
Note, that the indicative error bars in Table~\ref{t1} are not statistical, therefore in Eq.~(\ref{o1}) we 
present the simple average value. 

The CDW-induced broadening at Cu sites comes from contributions of $s$- and $s'$-waves, see Eq.~(\ref{anu1}).
Below we assume that the phases are locked, $\phi_s=\phi_{s'}$.
Hence comparing Eqs.~(\ref{anu}) and ~(\ref{anu1}) we conclude that for Cu sites
$A=94.3|A_{s}+0.23A_{s'}|$MHz.

Using the experimental widths presented in the top line of Table~\ref{t1} together with Eq.~(\ref{tlw})
we find
\begin{equation}
\label{cu1}
\left|A_{s}+0.23 A_{s'}\right|= 2.0 \cdot 10^{-3}\ .
\end{equation}
It is very natural to assume that the amplitudes $A_s$ and $A_{s'}$
are related as components of Zhang-Rice singlet, $A_s\approx 2A_{s'}$, see
Ref.~\cite{Haase2004}. Hence, using (\ref{o1}), (\ref{cu1}) we come to the following CDW amplitudes
at $T=60$K and $B\approx 12 - 15$T,
\begin{eqnarray}
\label{cdw1}
&&A_s=1.8 \cdot 10^{-3}, \ \ A_{s'}=0.87 \cdot 10^{-3},
\nonumber\\
&&A_d=3.8 \cdot 10^{-3}, \ \ \delta p=A_s+2A_{s'}=3.5 \cdot 10^{-3} \ .
\end{eqnarray}
The values of $A_{s,s',d}$ are in units of the number of holes per atomic site,
and $\delta p$ is the doping modulation  amplitude in units of the number of 
holes per unit cell of a CuO$_2$ plane.
Values of the amplitudes have not been reported previously,
but the ratios have been deduced from the polarization-resolved resonant X-ray scattering~\cite{Comin2015}.
Our ratio $A_{s'}/A_d\approx 0.23$ is reasonably close to the value $A_{s'}/A_d\approx 0.27$ obtained 
in Ref. \cite{Comin2015}, however the ratio $A_{s}/A_d\approx 0.47$ is significantly  larger
than the value reported in Ref. \cite{Comin2015}. 

Superficially our ratio $A_{s'}/A_d \approx 0.23$ is reasonably close to the value
$A_{s'}/A_d \approx 0.27$ obtained in Ref. \cite{Comin2015}, on the other hand the ratio
$A_s/A_d \approx 0.47$ is significantly larger than the value
reported in Ref. \cite{Comin2015}. 
However, one has to be careful making a direct comparison of our results with
Ref. \cite{Comin2015}.
The analysis \cite{Comin2015} assumes either $s + d$ or $s' + d$ models, while
we keep the three components ($s+s'+d$ model).
For example, it is easy to check that the $s'+d$ model ($s=0$) is inconsistent with
the NQR data, so the agreement in the value $A_{s'}/A_d \approx 0.27$ is purely accidental.
On the other hand, in principle we can fit the NQR data by the $s+d$ model ($s'=0$),
this results in $A_s/A_d \approx 0.53$ that is inconsistent with \cite{Comin2015}.

The ratios of the CDW amplitudes $A_s/A_d \sim 0.2$, $A_{s'}/A_d \sim 0.1$
have been reported in STM measurements with BSCCO and NaCCOC \cite{Fujita2014}. 
Comparing these ratios (although measured in different cuprates) with 
our results we see that \cite{Fujita2014} indicates dominance of the $d$-wave
component, whereas in our analysis the $s$-wave amplitude is
about a half of the $d$-wave amplitude.
We do not have an explanation for the discrepancy
between our results and REXS/STM measurements
\cite{Comin2015, Fujita2014}, moreover REXS and STM are inconsistent with each other.
The advantage of our analysis is that it is very simple
and straightforward and, of course, NQR is a bulk probe.

Unfortunately, NQR data for magnetic field $B\approx 30$T are not that
detailed as that for $B\approx 12 - 15$T. Nevertheless, based on the Cu
NQR/NMR line broadening measured in Ref.~\cite{Wu2011} and rectifying 
the Cu NQR line width (subtracting the Knight shift), we 
conclude that $\Gamma_0=\Gamma_{T=75K}=0.6$MHz and  $\Gamma_1=\Gamma_{T=1.3K}=1.0$MHz.  
Hence, using the same procedure as that for the low magnetic field, we find
$\left|A_{s}+0.23 A_{s'}\right|= 3.7 \cdot 10^{-3}$. 
Again, assuming the Zhang-Rice singlet ratio, $A_s\approx 2A_{s'}$,
we find the $s$-wave CDW amplitudes for $B\approx 30$T and $T=1.3$K:
\begin{eqnarray}
\label{cdw2}
&&A_s=3.3 \cdot 10^{-3}, \ \ A_{s'}=1.6 \cdot 10^{-3}, \nonumber\\
&&\delta p=A_s+2A_{s'}=6.5 \cdot 10^{-3} \ .
\end{eqnarray}
The doping modulation amplitude $\delta p$ is about two times smaller than the
estimate presented in Ref.~\cite{Wu2011}.
Unfortunately, data~\cite{Wu2013} are not sufficient for unambiguous subtraction 
of the Knight shift broadening from  oxygen lines, so the determination of $A_d$ is less 
accurate. However, roughly at $B\approx 30$T and $T=1.3$K the value is
$A_d\sim 6 \cdot 10^{-3}$.

To complete our phenomenological analysis we comment on two broad
classes of possible mechanisms of the CDW instability.
(i) In the first class  the CDW instability is driven purely by strongly 
correlated electrons which generate the charge wave. 
It can be due to electron-electron interaction mediated by spin fluctuations
and/or due to the  Coulomb interaction, 
see e.g. Refs~\cite{Volkov2015, Wang2014}. 
We call this class of CDW formation mechanisms 
the  ``electronic scenario''.
In this scenario phonons/lattice are not crucial for the CDW instability,
they are only spectators that follow electrons.
(ii) In the second class which we call the ``Peierls/Kohn scenario'' and which is
known in some other compounds~\cite{Renker,Johannes2008,Hoesch2009}, both
electrons and phonons are involved in the CDW development on equal footing.
We argue that the  phonon softening data might support the second class.

\begin{figure}[h] 
\includegraphics[scale=0.35]{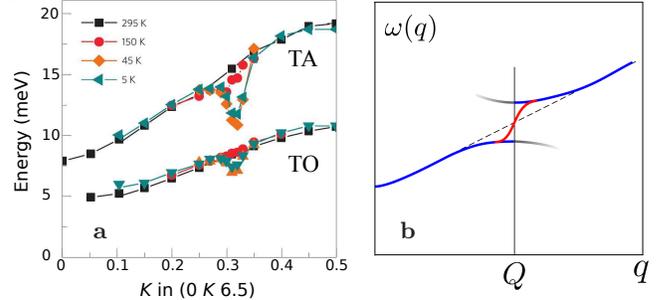}
\begin{picture}(0,0)
         \put(-213,23){\textbf{a}}
     \put(-97,23){\textbf{b}}
\end{picture}
\caption{\textbf{(a)}: Phonon anomaly (dip) in the dispersion of 
transverse acoustic (TA) and transverse optical (TO) modes at ${\bf q}=(0,Q,6.5)$ 
in YBa$_2$Cu$_{3}$O$_{6.6}$~\cite{LeTacon2014}.  
\textbf{(b)} The expected phonon dispersion in the ``electronic 
scenario''. Blue solid lines show the dispersion in a perfect long-range CDW,
grey lines represent the shadow bands. For a finite CDW correlation length
the shadow bands practically disappear and solid blue lines become 
connected by the red solid line.
}
\label{F5}
\end{figure}

A very significant softening of transverse acoustic and transverse optical  
modes in YBCO has been observed in Ref.~\cite{LeTacon2014}.
The softening data are reproduced in Fig.~\ref{F5}a.
The anomaly is very narrow in momentum space, $\delta q \approx 0.04$ r.l.u.,
and it is only two times broader than  the width of the elastic CDW peak,
$\delta q = 1/\xi_{a,b} \approx 0.02$ r.l.u. measured in RIXS and 
XRD~\cite{Ghiringhelli2012, Chang2012, Blackburn2013}.
Our observation is very simple, in the case of the ``electronic'' scenario,
the electronic CDW creates a weak periodic potential 
$V=V_0 \cos({\bf Q}\cdot{\bf r} + \phi)$ for phonons.
Diffraction of phonons from the potential must lead to the usual
band-structure discontinuity of the phonon dispersion $\omega_q$ at
$q=Q$ as it is shown by blue lines in Panel b of Fig.~\ref{F5}.
In the presence of the finite correlation length of the CDW the discontinuity is healed as it is
shown by the red solid line at the same panel, and combined blue-red solid line
shows the expected phonon dispersion. 
In Supplementary material we present a calculation which supports this 
picture, but generally the picture is very intuitive.
Obviously, this physical picture for the phonon dispersion
is qualitatively different from the experimental data in Panel a of Fig.~\ref{F5}.
On the other hand the ``Peierls/Kohn scenario'', where both
electrons and phonons are equally involved in the CDW development, 
leads to phonon dispersions like that in Fig.~\ref{F5}a. This has been
observed in several compounds, see e.g. Refs.~\cite{Renker,Hoesch2009}.
Even though
the phenomenological observation does not explain the
mechanism of the CDW in underdoped cuprates, the observation
poses a significant challenge to theoretical models
based on ``the electronic scenario'' of the CDW formation.
Furthermore, the phonon softening is generally expected
in the ``Kohn/Peierls scenario'', which is likely to be the case in YBCO. At the same time
our arguments in favour of the ``Kohn/Peierls
scenario'' are not quite conclusive. 
Indeed, it seems that the ``Kohn/Peierls scenario'' does not provide an explanation
of the strong broadening of TA and TO phonon
modes at $T<T_{CDW}$ [19], as well as it does not explain
why the phonon softening appears only in superconducting
state $T<T_c$. 
So the last point of our work is less solid than the main
results concerning the amplitudes of the CDW.
Nevertheless, we believe that the presented discussion of 
the ``Kohn/Peierls scenario'' versus ``electronic scenario''
is useful for future work on the microscopic mechanism of the CDW.

In conclusion, our  analysis of available experimental data has resolved
open problems in the phenomenology of the charge density wave (CDW) in underdoped cuprates.
We have determined the amplitudes of $s$-, $s'$-, and $d$-wave components of the
density wave. The amplitudes at low magnetic field and temperature 
$T=60$K are given in Eq.~(\ref{cdw1}), and the amplitudes for 
magnetic field $B=30$T and temperature $T=1.3$K
are given in Eq.~(\ref{cdw2}).
We show that the data rule out a checkerboard pattern, and we also argue
that the data might rule out mechanisms of the CDW which do not include phonons.


\subsection{Acknowledgements}
We thank M.-H. Julien, M. Le Tacon, J. Haase and A. Damascelli for communications, discussions, and
very important comments. The work has been supported by the Australian Research Council, 
grants DP110102123 and DP160103630.

\subsection{Author contribution}
Both authors, Y. A. K. and O. P. S,  contributed to all aspects of this work.

\subsection{Competing financial interests}
 The authors declare no competing financial interests.

\newpage

\appendix

\section{Ginzburg-Landau model of the CDW and phonon dispersion}
Here we consider a simple phenomenological Ginzburg-Landau model of the CDW
developed in purely electron sector. 
Phonons weakly interact with electrons
and follow the developed CDW as ``spectators''.
This is the first scenario discussed in the main text.
Our analysis below shows that the phonon softening data does not support this scenario.

\begin{figure}[h]
\vspace{5pt}
\includegraphics[scale=0.4]{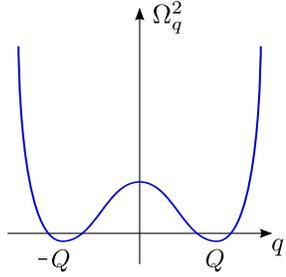}
\caption{A sketch of the effective kinetic energy  $\Omega^2_q$ of the 
electronic CDW mode $\psi$. 
 }\label{FO}
\end{figure}

Let us consider a quasi-one dimensional (stripe-like) CDW, we direct the $x$ 
axis along the  CDW wave-vector. The CDW can be represented as a collective bosonic mode ($\psi$) in the electronic system. 
The  effective Ginzburg-Landau-like Lagrangian for the CDW mode $\psi$ reads
\begin{equation}\label{eq:L_psi}
\mathcal{L}_\psi = \frac{\dot\psi^2}{2} - \psi\frac{\hat\Omega^2}{2} \psi - \frac{\alpha}{4} \psi^4 \ ,
\end{equation}
where $\psi(r)$ is a variation of electron density, $\hat\Omega^2/2$ is operator of 
``stiffness'' of the CDW mode, and $\alpha > 0$ is a  self-action constant.
In momentum representation  $\Omega^2_q$ is a simple function sketched in 
Fig. \ref{FO}.
Kinetic energy has minima  at $q_x= \pm Q$, and we can represent it as
\begin{equation}
\Omega_q^2 = \Omega_Q^2 + \kappa (q_x^2 - Q^2)^2/4,
\end{equation}
where $\kappa$ is a some constant.
Importantly the minimal value of  $\Omega_{\pm Q}^2$ is negative, $\Omega_{\pm Q}^2 < 0,$
providing formation of the incommensurate CDW with the wave vector $Q$.
The density variation is real, hence 
\begin{equation}\label{eq:psi_Q}
\psi(x) = 2|\psi_Q|\cos(Qx+\phi),
\end{equation}
here $\psi_Q$ is a Fourier component of $\psi(r)$.
In a perfect system the phase $\phi$ is arbitrary and this must result
in a Goldstone sliding mode. Of course a disorder pins the phase.
A similar phenomenological approach was successfully applied to describe a CDW 
state in transition-metal dichalcogenides \cite{McMillan1975}.

To find the CDW amplitude one has to minimize the energy
\begin{equation}\label{eq:energy}
E_\psi = \int d^2 r \left( \psi \frac{\hat\Omega^2}{2} \psi + \frac{\alpha}{4} \psi^4\right).
\end{equation} 
The saddle-point equation for static $\psi(r)$ reads
\begin{eqnarray}\label{eq:saddle_p}
\hat\Omega^2\psi + \alpha \psi^3 = 0.
\end{eqnarray}
Performing Fourier transform in Eqs. (\ref{eq:saddle_p}) and leaving only 
the dominating Fourier components with $q=\pm Q$ we find
\begin{equation}
\label{psiq}
|\psi_Q|^2 =  -\Omega^2_Q/3\alpha. 
\end{equation}

To find the CDW excitation spectrum we expand energy (\ref{eq:energy}) up 
to second order in fluctuations on the top of the  ground state 
(\ref{eq:psi_Q}),(\ref{psiq}),
\begin{equation}\label{eq:delta^2E}
\delta^{2}E_\psi = \int d^2 r \left( 
 \delta\psi \frac{\hat\Omega^2}{2} \delta\psi + \frac{3 \alpha}{2}  \psi^2 (\delta\psi)^2 \right).
\end{equation}
The term $\propto \psi^2  = |\psi_Q|^2(2+e^{2i(Qx+\phi)} + e^{-2i(Qx+\phi)})$ 
in (\ref{eq:delta^2E}) plays role of the effective potential with wave vector 
$2Q$ for the $\psi$-excitations. This results in a mixing 
between $\delta\psi_q$ and $\delta\psi_{q-2Q}$ (hereafter 
we assume that $q > 0$). 
Therefore, it is convenient to write the excitation energy as
\begin{equation}
\label{eq:Hamilt_eff}
\delta^2 E_\psi = 
\sum_{q} \Psi_q^\dag
\begin{pmatrix}
\Omega^2_q + 2|\Omega^2_Q| & |\Omega^2_Q|e^{2i\phi} \\
|\Omega^2_Q|e^{-2i\phi}  & \Omega^2_{q-2Q} + 2|\Omega^2_Q| & 
\end{pmatrix}\Psi_q  ,
\end{equation}
where
\begin{equation}
\Psi_q=
\left(
\begin{array}{l}
\delta\psi_q\\
\delta\psi_{q-2Q}
\end{array}
\right).\nonumber
\end{equation}
Euler-Lagrange equation corresponding to  (\ref{eq:L_psi}), (\ref{eq:Hamilt_eff}) results in the two normal
modes: the sliding Goldstone mode and the gapped Higgs mode with the energies
\begin{eqnarray}
&&\epsilon_{Gq} = \sqrt{\kappa Q^2}|p_x|,\nonumber\\
&&\epsilon_{Hq} = \sqrt{2|\Omega^2_Q| + \kappa Q^2p_x^2}\nonumber\\
&&\bm p= \bm q- \bm Q \ .
\end{eqnarray}
The corresponding eigenmodes are
\begin{equation}
\begin{array}{c}
\label{mix}
G_q = 1/\sqrt{2}(e^{i\phi}\delta\psi_q-e^{-i\phi}\delta\psi_{q-2Q}),\\
H_q = 1/\sqrt{2}(e^{i\phi}\delta\psi_q+e^{-i\phi}\delta\psi_{q-2Q}).
\end{array}
\end{equation}
Interestingly, due to the parabolic behaviour of $\Omega^2_q$
near $q_x=\pm Q$, see Fig. \ref{FO}, the weights of the states with wave vectors $q$ and $q-2Q$ in (\ref{mix}) are equal in a broad range of momenta around $q=\pm Q$.

\begin{figure}[h]
\vspace{5pt}
\includegraphics[scale=0.25]{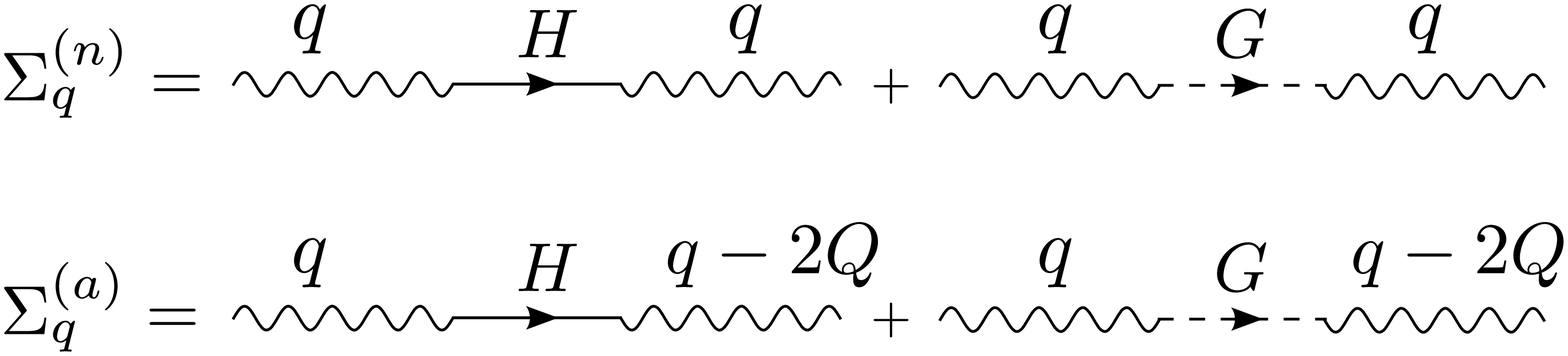}
\caption{Normal and anomalous phonon self-energy
 }\label{FS}
\end{figure}

The phonon is described by field $\varphi$, the phonon Lagrangian
reads
\begin{equation}
\mathcal L_{\varphi} = \frac{\dot\varphi^2}{2} -  \varphi \frac{\hat\omega^2}{2} \varphi,
\end{equation}
where $\hat\omega^2/2$ is the elastic energy of lattice deformation.
In momentum representation it is equivalent to
 the bare phonon dispersion $\omega_q$.
The CDW and phonons weakly interact, we describe the interaction by the
Lagrangian
\begin{equation}
\label{Li}
\mathcal L_{int} = -  \lambda \psi \varphi \ ,
\end{equation}
where $\lambda$ is coupling constant.
Due to the coupling the CDW creates phonon condensate at $q_x=\pm Q$ 
(static lattice deformation) with amplitude $\varphi_Q = -\lambda/\omega^2_Q\, \psi_{-Q}$.
Let us now consider phonon dispersion in the presence of the collective CDW mode.
The interaction (\ref{Li}) in combination with Eqs (\ref{mix}) results
in the following vertexes describing transition of phonon to the Higgs and 
Goldstone modes of the CDW.
\begin{eqnarray}
\label{vert}
&&\langle\delta\varphi_q| H_q\rangle= \langle \delta\varphi_q|G_q\rangle=
\frac{\lambda}{\sqrt{2}} e^{i\phi}
\nonumber\\
&&\langle \delta\varphi_{q-2Q}|H_q\rangle= 
-\langle \delta\varphi_{q-2Q}|G_q\rangle=
\frac{\lambda}{\sqrt{2}} e^{-i\phi} \ .
\nonumber
\end{eqnarray}
This leads to normal and anomalous phonon self-energy operators
shown in Fig. \ref{FS}.
The corresponding analytical expressions are
\begin{eqnarray}
\label{sig}
&&\Sigma^{(n)}_{q} = \frac{\lambda^2}{2} \left( \frac{1}
{\omega_q^2 - \epsilon_{Hq}^2 } + 
\frac{1}{\omega_q^2 - \epsilon_{Gq}^2 }\right),\nonumber \\
&&\Sigma_{q}^{(a)} = e^{2i\phi}\Delta_q^2,\nonumber\\
&&\Delta_q^2=\frac{\lambda^2}{2} 
\left( \frac{1}
{\omega_q^2 - \epsilon_{Hq}^2 } - 
\frac{1}{\omega_q^2 - \epsilon_{Gq}^2 }\right).
\end{eqnarray}
A selfenergy generally depends on $q$ and $\omega$. In Eqs.(\ref{sig})
we set $\omega=\omega_q$.
So ''renormalized'' dispersion $\widetilde\omega_q$ of phonons is described by the eigenvalue
problem
\begin{equation}
\label{d22}
\widetilde\omega_q^2\Phi_q  =
\begin{pmatrix}
\omega_q^2+\Sigma^{(n)}_{q} & e^{2i\phi}\Delta_q^2 \\
e^{-2i\phi}\Delta_q^{2}& \omega_{q-2Q}^2+\Sigma^{(n)}_{q-2Q}
\end{pmatrix}\Phi_q \ ,
\end{equation}
where
\begin{equation}
\Phi_q=
\left(
\begin{array}{l}
\delta\varphi_q\\
\delta\varphi_{q-2Q}
\end{array}
\right).
\end{equation}
As it is intuitively clear even without a calculation Eq. (\ref{d22})
is equivalent to scattering of phonon from some effective periodic
potential with wave vector $2Q$.

One can represent matrix elements in (\ref{d22}) as
\begin{eqnarray}
\label{lin}
&&\sqrt{\omega_q^2+\Sigma^{(n)}_{q}}\approx w+cp_x,\nonumber\\
&&\sqrt{\omega_{q-2Q}^2+\Sigma^{(n)}_{q-2Q}}\approx w-cp_x,\nonumber\\
&&\Delta_q\approx\Delta \ ,
\end{eqnarray}
when the detuning $\bm p= \bm q- \bm Q$ is small. The speed $c$ follows from the overall
slope of the phonon dispersion.
For the TO mode $w\approx 15$ meV, $c\approx 30$ meV/r.l.u.,
see Fig. 5a in the main text.
The phonon dispersion which follows from (\ref{d22}) and (\ref{lin}),
\begin{eqnarray}
\label{di}
\widetilde\omega_q^2\approx w^2\pm\sqrt{4 w^2 c^2 p_x^2+\Delta^4} \ ,
\end{eqnarray}
is shown in Fig.5b in the main text by the blue solid line.
Shadow bands are indicated  by fading grey  lines.
Expected intensity of the shadow bands is extremely small.
Intensities of the bright (br) and shadow (sh) modes are
\begin{eqnarray}
&& I_{br} \propto |\delta \varphi_q|^2=
\frac{1}{2}\left(1+\frac{2 w c p_x}{\sqrt{4 w^2 c^2 p_x^2 +\Delta^2}}\right),
    \nonumber\\ 
&& I_{sh} \propto |\delta \varphi_{q-2Q}|^2=
\frac{1}{2}\left(1-\frac{2 w c p_x}{\sqrt{4 w^2 c^2 p_x^2 +\Delta^2}}\right).
   \nonumber
\end{eqnarray}
If the gap in the phonon spectrum ($gap \approx \Delta^2/w$) is 3 meV (TO mode),
the intensity of the shadow mode practically diminishes at detuning 
$p_x= q_x-Q \geq 0.04$ r.l.u.

So far we disregarded temperature and intrinsic disorder.
We remind the following  experimental observations: 
(i) the CDW onset temperature is $T_{CDW}\sim 150$K, 
(ii) the CDW in low/zero magnetic fields is essentially two-dimensional, the correlation length
in the $c$-direction is about one lattice spacing while the in-plane
correlation length is $\xi_{a,b} \sim 20$  lattice spacings. 
In agreement with Mermin-Wagner theorem the observation (ii) implies
that onset of the CDW at $T = T_{CDW}$ is not a true phase transition,
it is a two-dimensional freezing crossover.
Hence at $T > T_{CDW}$ the phase $\phi$  fluctuates
with time and as a result the off-diagonal matrix element in
Eq. (\ref{d22}) is averaged to zero, $e^{2i\phi}\Delta_q^2\to 0$.
Hence the phonon dispersion at $T > T_{CDW}$ near $q_x=+Q$ is
\begin{eqnarray}
\widetilde\omega_q=\sqrt{\omega_q^2+\Sigma^{(n)}_{q}}\approx w+cp_x \ .
\end{eqnarray}
At $T < T_{CDW}$ the temporal fluctuations freeze, however due to the quenched
disorder the phase $\phi(r)$ is a fluctuating function of coordinate $r$
with the correlation length $\xi_{a,b}$. If $|p_x|> 1/\xi_{a,b}=0.02$ r.l.u. the spatial fluctuations
are not relevant and the phonon dispersion is given by Eq.(\ref{di}).
However, if the detuning is small, $|p_x| \ll 1/\xi_{a,b}$, one must average
over spatial fluctuations of the phase $\phi$ effectively vanishing
the off-diagonal matrix element, $e^{2i\phi}\Delta_q^2\to 0$.
This results in the red solid line connecting the blue solid lines in
Fig. 5b  in the main text.

All in all the  phonon dispersion expected if the ``electronic'' scenario
is realized is shown in Fig. 5b (main text).
Obviously, the dispersion is inconsistent with the data.
Therefore, we rule out the ``electronic'' scenario.

\end{document}